\documentclass[DIV=13, a4paper]{scrartcl}

\usepackage[utf8]{inputenc}
\usepackage{empheq}
\usepackage{amssymb}
\usepackage{cite}
\usepackage{mathbbol}
\usepackage{latexsym}
\usepackage{ifpdf}
\usepackage{esint}
\usepackage{slashed}
\usepackage{color}

\usepackage{tikz}
\newcommand*\circled[1]{\tikz[baseline=(char.base)]{\node[shape=circle,draw,inner sep=0.3pt] (char) {#1};}}

\newcommand{\bS}{\boldsymbol{\mathcal{S}}}
\newcommand{\bD}{\boldsymbol{\mathcal{D}}}

\begin{document}
\begin{center}
\begin{Large}
 \textbf{\sffamily Accounting for dissipation in the scattering approach\\[0.2\baselineskip]
                         to the Casimir energy}
\end{Large}

\vspace{\baselineskip}
\begin{large}
Romain Guérout\,$^{1}$\footnote{guerout@lkb.upmc.fr},
Gert-Ludwig Ingold\,$^{2}$,
Astrid Lambrecht\,$^{1}$ and
Serge Reynaud\,$^{1}$
\end{large}

$^{1}$\,Laboratoire Kastler Brossel (LKB), UPMC-Sorbonne Universit\'e, CNRS, ENS-PSL Research University,
Coll\`ege de France, F-75252, Paris, France\\
$^{2}$\,Universität Augsburg, Institut für Physik, D-86135 Augsburg, Germany
\end{center}

\begin{abstract}
We take dissipation into account in the derivation of the Casimir energy
formula between two objects placed in a surrounding medium. The dissipation
channels are considered explicitly in order to take advantage of the unitarity
of the full scattering processes. We demonstrate that the Casimir energy is given
by a scattering formula expressed in terms of the scattering amplitudes coupling
internal channels and taking dissipation into account in an implicit way.
We prove that this formula is also valid when the surrounding medium is dissipative.
\end{abstract}

\section{Introduction} 
\label{sec:introduction}

In the last decades, Casimir physics has known a renewed interest thanks to
new measurements of the Casimir interaction between macroscopic 
objects~\cite{casimir1948attraction,lifshitz1956theory,schwinger1978casimir} 
with an improved precision~\cite{decca2005precise,decca2007tests,masuda2009limits,%
sushkov2011new,chang2012gradient} as well as efforts to meet the 
associated theoretical challenges~\cite{lambrecht2000casimir,bostrom2000thermal,%
milton2009recent,rahi2009scattering,klimchitskaya2009casimir,ingold2009quantum}.
In order to accurately reproduce the experimental data, a theoretical
calculation has to model the optical properties of the materials used.
A puzzling result of these comparisons is that some of the most precise 
experiments appear to agree well with the calculations only when the Ohmic
losses in the metallic plates are neglected in the model. 
Several possible explanations of this puzzle have been discussed
but none of them seem to be satisfactory
(a recent review is presented in~\cite{reynaud2017casimir}). 
For example the electrostatic interaction between patches on the 
plates is certainly a possible systematic effect for Casimir force
measurements~\cite{speake2003forces,kim2010surface,behunin2012modeling}
but it does not explain the discrepancy between theory and 
measurements~\cite{behunin2012electrostatic,behunin2014kelvin}. 

This yet unsolved discrepancy between experiment and theory has
spearheaded discussions about the correctness of the theoretical formula
used to describe Casimir interaction. In particular, it has been recently
realized~\cite{intravaia2009casimir,guerout2014derivation,guerout2016lifshitz} 
that the calculations using the lossless plasma model were in fact neglecting 
the interaction between magnetically coupled induced currents due to a subtlety in the
mathematical description of causality properties of the metallic optical response. 
Though it does not solve the discrepancy, this work has shed interesting new light
on the derivation of the scattering formula used in most calculations.
Among other worries, it has also been suggested that the scattering might not be 
valid for the dissipative metallic plates used in the experiments~\cite{bordag2011drude}.
Some works have been devoted to \emph{ab initio} treatments of the
Casimir interaction between dissipative  
mirrors~\cite{jancovici2005casimir,buenzli2005casimir,intravaia2012casimir,bordag2017casimir}. 

In the present article, we show that dissipation is taken into account in the
usual scattering formula of the Casimir interaction
energy~\cite{jaekel1991casimir,ingold2015casimir}.  We consider explicitly the
channels responsible for dissipation in order to take advantage of the
unitarity of the scattering processes.  In the end, the Casimir energy is given
by a scattering formula written in terms of the scattering amplitudes of the
mirrors, taking into account in an implicit way the channels responsible for
dissipation. In the context of Casimir physics, this result was already proven
for the particular case of the plane-plane
geometry~\cite{genet2003casimir,lambrecht2006casimir}, and the derivation in
the present paper can be considered as a generalization to the case of an
arbitrary geometry. In a broader context, it is reminiscent of properties known
in the theory of resistance in mesoscopic physics~\cite{engquist1981definition}
or that of quantum field propagation in a dissipative
medium~\cite{huttner1992quantization,matloob1995electromagnetic}.


\section{Scattering interpretation of the Casimir effect} 
\label{sec:scattering_interpretation_of_the_casimir_effect}

Our starting point is the interpretation of the Casimir effect in terms of the
scattering formula~\cite{jaekel1991casimir,ingold2015casimir}. Since
temperature does not play a key role in the considerations presented below, we
assume $T=0$ for the sake of simplicity.  We begin by considering a single
object placed into a medium, with scattering of electromagnetic fluctuations by
this object leading to a change of the vacuum energy written in terms of its
scattering matrix $\textbf{S}$
\begin{equation}
	\label{eq:vacuumEnergy}
	\Delta E_{\text{vac}} = -\hbar \int_{0}^{\infty} \frac{d \omega}{2 \pi} \Delta\phi\,,
\end{equation}
\begin{equation}
	\label{eq:scatteringPhase}
	\Delta\phi = \frac{1}{2 i} \log \det \textbf{S}\,.
\end{equation}
The change of vacuum energy $\Delta E_\text{vac}$ is infinite when calculated
for a single object, but its part relevant for estimating the Casimir effect
turns out to be
finite~\cite{schwinger1954theory,balian1977electromagnetic,plunien1986casimir}.
The phase shift $\Delta\phi$ is the trace of eigen-phase shifts summed over all
scattering channels at a given frequency $\omega$. The formula
(\ref{eq:vacuumEnergy}) thus has a clear physical meaning when the scattering
matrix $\textbf{S}$ is unitary, as it should if all scattering channels are
taken into account. Accordingly, it is obvious that $\Delta E_{\text{vac}}$ is
real.

This discussion does not mean that \eqref{eq:vacuumEnergy} cannot be applied
when dissipation enters the game. It only implies that all scattering channels
responsible for dissipation processes must be included in the scattering
theory. This can always be achieved and necessarily leads to a unitary
matrix. The general expression \eqref{eq:vacuumEnergy} always describes the
modification of the vacuum energy due to the presence of scatterers. Another
way to see that is to transform equation \eqref{eq:vacuumEnergy} into an
equivalent equation through an integration by parts and a rearrangement of
terms
\begin{equation}
	\label{eq:vacuumEnergy2}
	\Delta E_{\text{vac}} = \int_{0}^{\infty} d \omega \frac{\hbar \omega}{2} \Delta \eta\,,
\end{equation}
\begin{equation}
	\label{eq:deltaDOS}
        \Delta \eta = \frac{1}{\pi} \frac{\partial}{\partial \omega} \Delta\phi\,.
\end{equation}
Here, $\hbar\omega/2$ describes the vacuum energy of one mode at frequency $\omega$, 
while $\Delta \eta$ is the modification of the density of states due to the presence 
of the scatterer~\cite{plunien1986casimir,souma2002local}. Here again, this interpretation 
of \eqref{eq:deltaDOS} has a direct physical meaning when the scattering matrix is unitary. 

\begin{figure}
	\centering
	\includegraphics[width=0.6\textwidth]{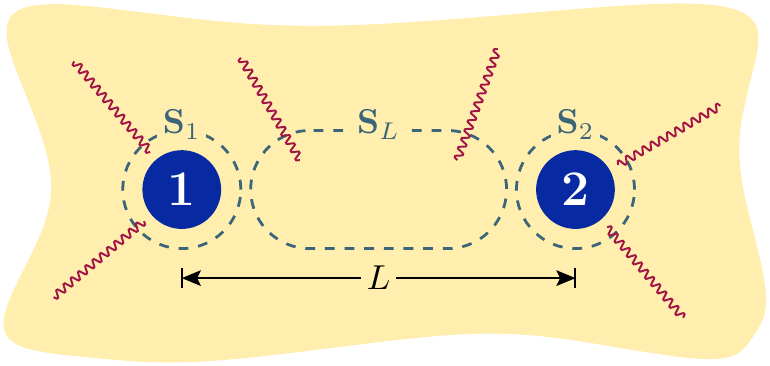}
	\caption{The Casimir interaction between two objects \protect\circled{1} and
	\protect\circled{2} at a distance $L$ is considered. As indicated by the
	wavy lines, both objects as well as the medium in between are in
	general dissipative. The two objects are described by unitary scattering
	matrices $\mathbf{S}_1$ and $\mathbf{S}_2$ which account also for the
	external channels associated with the dissipation. The unitary scattering
	matrix $\mathbf{S}_L$ describes the translation between the reference
	frames of objects \protect\circled{1} and \protect\circled{2} and also accounts for
	the external channels.}
	\label{fig:casimirSetup}
\end{figure}

In the following we derive the expression for the Casimir interaction energy
between two objects \circled{1} and \circled{2}. The set-up is displayed in
Figure~\ref{fig:casimirSetup} with wavy lines representing dissipative channels
for the objects and the medium.  We apply the formula written above for the
total scattering matrix \textbf{S} viewed as describing the change of the
electromagnetic vacuum energy when two objects are placed in the surrounding
medium at a distance $L$. As depicted in Figure~\ref{fig:casimirSetup}, the
total scattering matrix \textbf{S} can be decomposed into the scattering
matrices $\textbf{S}_1$ and $\textbf{S}_2$ related to the individual objects
and the matrix $\textbf{S}_L$ describing the propagation between the two
objects over a distance $L$ through the medium. The expression for the Casimir
interaction energy is then obtained as the change in the vacuum energy caused
by the full scattering matrix \textbf{S} after extracting the part depending on
the distance $L$.

We now introduce the notion of \emph{internal} and \emph{external} scattering
channels. An internal scattering channel links the two objects. It represents,
for example, an outgoing channel from object \circled{1} which becomes an
incoming channel at object \circled{2} after propagation by a translation
matrix as discussed in Section~\ref{sec:casimirEnergy}. The channels which are
not internal are named external channels. They include all channels responsible
for dissipation processes. Once these channels are included, the scattering
matrices $\textbf{S}_1$, $\textbf{S}_2$ and $\textbf{S}_L$ are unitary, and
therefore the total scattering matrix $\textbf{S}$ is unitary as well. The
Casimir interaction is then given by the part of eqs.~\eqref{eq:vacuumEnergy}
and \eqref{eq:vacuumEnergy2} which depends on $L$. We show below that the
Casimir energy can also be described by a simplified scattering formula written
in terms of scattering amplitudes between internal channels only, with the
channels responsible for dissipation taken into account in an implicit
way~\cite{genet2003casimir,lambrecht2006casimir}.

\section{Determinant formula for two scatterers}
\label{sec:detFormula}

\begin{figure}
	\centering
	\includegraphics[width=0.6\textwidth]{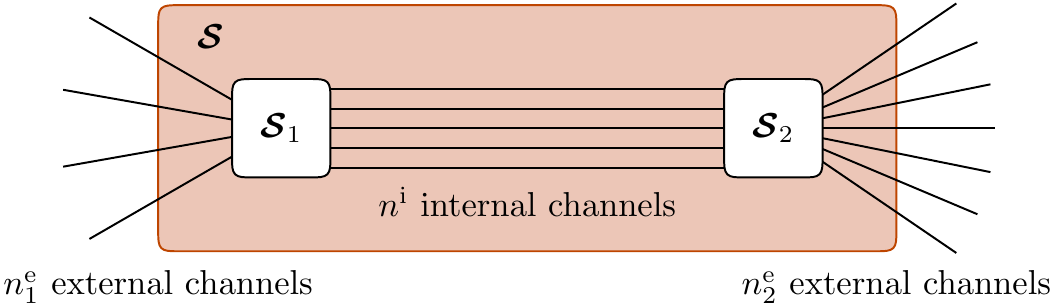}
	\caption{Scattering geometry with internal structure. Seen from the
	outside, a total of $n^\text{e}_1+n^\text{e}_2$ external channels are coupled by a
	scattering matrix $\bS$. The internal structure is accounted for
	by two scattering matrices $\bS_1$ and $\bS_2$ coupling
	$n^\text{i}$ internal channels to $n^\text{e}_1$ and $n^\text{e}_2$ external channels,
        respectively.}
        \label{fig:generalSetup}
\end{figure}

In this section, we derive a relation involving determinants of scattering
matrices for a scattering set-up with an internal structure described by two
scattering matrices as depicted in Figure~\ref{fig:generalSetup}. In order to
emphasize, that the involved scattering matrices are general and not
necessarily related to the scattering matrices introduced in
Figure~\ref{fig:casimirSetup}, we denote them by calligraphic symbols $\bS$,
$\bS_1$, and $\bS_2$. When applying the relation for the determinant
(\ref{eq:detFinal}) obtained at the end of this section, we will replace these
general scattering matrices by specific scattering matrices related to the
set-up shown in Figure~\ref{fig:casimirSetup}.

Ignoring the internal structure, the scattering properties can be described by
a scattering matrix $\bS$ coupling the $n^\text{e}_1+n^\text{e}_2$
external channels among each other. Accounting for the internal structure, in
addition to the $n^\text{e}_1$ and $n^\text{e}_2$ external channels associated
with the scattering matrices $\bS_1$ and $\bS_2$, respectively, one has $n^\text{i}$
internal channels coupling the two scatterers. Even though the two scatterers in
Figure~\ref{fig:generalSetup} are drawn at a certain distance, for the purpose
of this section we do not imply any effects of translation between the two
scatterers. Such effects can be accounted for by an additional
scattering matrix as we will see in Section~\ref{sec:casimirEnergy}.

\begin{figure}
	\centering
	\includegraphics[width=0.8\textwidth]{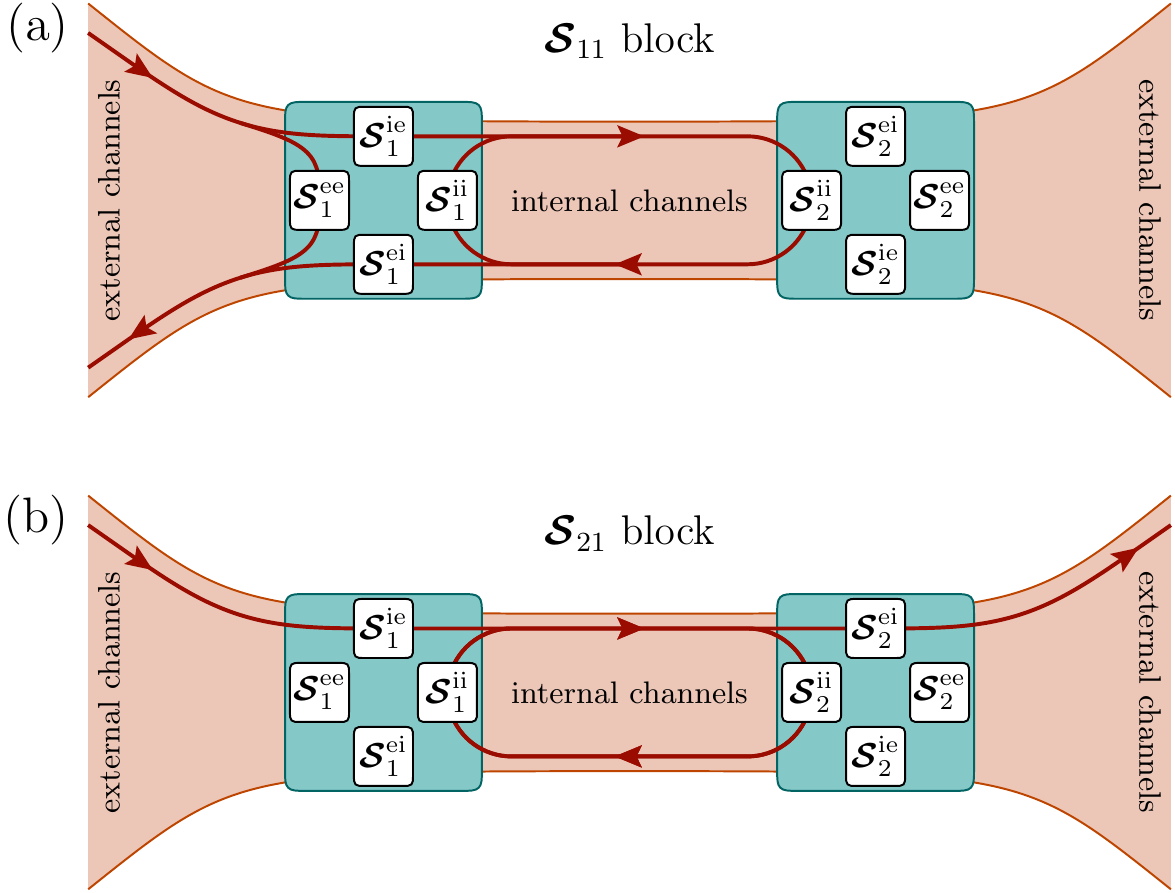}
	\caption{Schematic representation of the blocks (a) $\bS_{11}$ and
	(b) $\bS_{21}$ of the total scattering matrix $\bS = \bS_{1} \star \bS_{2}$.
	The diagrams visualize the equations \eqref{eq:blockOfS1} and \eqref{eq:blockOfS3},
	respectively. The two other blocks defined in \eqref{eq:blockOfS2} and
	\eqref{eq:blockOfS4} are obtained by exchanging the two objects.}
	\label{fig:scats}
\end{figure}

As the individual scattering matrices $\bS_1$ and $\bS_2$ couple
internal (i) and external (e) channels among each other, we can express them in block
matrix form as
\begin{equation}
	\label{eq:blockS1}
	\bS_k = \begin{pmatrix}
		\bS_k^\text{ii} & \bS_k^\text{ie} \\
		\bS_k^\text{ei} & \bS_k^\text{ee}
	\end{pmatrix}\qquad k=1, 2\,.
\end{equation}
The global scattering matrix $\bS$ is obtained by chaining the effect
of the two individual scatterers
\begin{equation}
	\label{eq:calSk}
	\bS = \bS_1 \star \bS_2
\end{equation}
where the symbol $\star$ indicates that $\bS$ is not obtained by a simple
matrix multiplication of $\bS_1$ and $\bS_2$. In fact, the scattering matrices
can be transformed into transfer matrices for which the chaining corresponds to
a matrix multiplication \cite{genet2003casimir}. From the resulting transfer
matrix, one obtains the global scattering matrix which can be expressed as a
block matrix
\begin{equation}
	\label{eq:blockS}
	\bS = \begin{pmatrix}
		\bS_{11} & \bS_{12} \\
		\bS_{21} & \bS_{22}
	\end{pmatrix}
\end{equation}
where the blocks refer to the external channels associated with scatterers 1 and 2.
Evaluating the chaining operation on $\bS_1$ and $\bS_2$ as just
described, one finds
\begin{subequations}
	\label{eq:blockOfS}
	\begin{align}
		\bS_{11} &= \bS_1^\text{ee} + \bS_1^\text{ei}\bS_{2}^\text{ii}\bD_{21}\bS_1^\text{ie}%
		            \label{eq:blockOfS1} \\
		\bS_{12} &= \bS_1^\text{ei}\bD_{12}\bS_2^\text{ie}\label{eq:blockOfS2} \\
		\bS_{21} &= \bS_2^\text{ei}\bD_{21}\bS_1^\text{ie}\label{eq:blockOfS3} \\
		\bS_{22} &= \bS_2^\text{ee} + \bS_2^\text{ei}\bS_1^\text{ii}\bD_{12}\bS_2^\text{ie}%
		                   \label{eq:blockOfS4}
	\end{align}
\end{subequations}
where
\begin{subequations}
	\label{eq:d}
	\begin{align}
		\bD_{12} &= \left(\mathbf{1}-\bS_2^\text{ii}\bS_1^\text{ii}\right)^{-1}\label{eq:d12}\\
		\bD_{21} &= \left(\mathbf{1}-\bS_1^\text{ii}\bS_2^\text{ii}\right)^{-1}\label{eq:d21}\,.
	\end{align}
\end{subequations}
The matrices \eqref{eq:d} account for an arbitrary number of round trips along the internal channels
between the two scatterers starting on scatterer 1 and scatterer 2,
respectively, as can be seen by means of a Taylor expansion, e.g.
\begin{equation}
	\bD_{12} = \mathbf{1}+\bS_2^\text{ii}\bS_1^\text{ii}+\bS_2^\text{ii}\bS_1^\text{ii}\bS_2^\text{ii}\bS_1^\text{ii}
	           +\bS_2^\text{ii}\bS_1^\text{ii}\bS_2^\text{ii}\bS_1^\text{ii}\bS_2^\text{ii}\bS_1^\text{ii}+\cdots
\end{equation}
The relations \eqref{eq:blockOfS1} and \eqref{eq:blockOfS3} are
visualized in Figure~\ref{fig:scats} and the other relations are obtained by
interchanging the two scatterers.

The relations \eqref{eq:blockS}, \eqref{eq:blockOfS}, and \eqref{eq:d} allow us
now to determine the determinant of the scattering matrix $\bS$. In the
derivation, we suppose that the three matrices $\bS$,
$\bS_1$, and $\bS_2$ are unitary. From the property
\eqref{eq:app4} of the determinant of a unitary $2 \times 2$ block matrix, we
get together with the relations \eqref{eq:blockOfS1} and \eqref{eq:blockOfS4}
\begin{equation}
	\label{eq:det1}
	\det \bS = \frac{\det(\bS_{22})}{\det(\bS_{11}^{\dagger})}
	         = \frac{\det(\bS_2^\text{ee} + \bS_2^\text{ei}\bS_1^\text{ii}\bD_{12}\bS_2^\text{ie})}%
			{\det(\bS_1^\text{ee} + \bS_1^\text{ei}\bS_{2}^\text{ii}\bD_{21}\bS_1^\text{ie})^{*}}\,.
\end{equation}
Then, we use a generalization of the matrix determinant lemma on the above
expression (see the Appendix). For instance, for the numerator we have according to
\eqref{eq:app6}
\begin{equation}
	\label{eq:det2}
	\det(\bS_2^\text{ee} + \bS_2^\text{ei}\bS_1^\text{ii}\bD_{12}\bS_2^\text{ie}) =
	\det(\bS_2^\text{ee}) \det(\bD_{12})
	\det(\bD_{12}^{-1} + \bS_2^\text{ie}{\bS_2^\text{ee}}^{-1}\bS_2^\text{ei}\bS_1^\text{ii})\,.
\end{equation}
By applying \eqref{eq:app4} to the matrices $\bS_1$ and
$\bS_2$, we can express the determinants of the blocks
$\bS^\text{ee}_1$ and $\bS^\text{ee}_2$ related to the external
channels by those related to the internal channels, $\bS^\text{ii}_1$
and $\bS^\text{ii}_2$, and obtain
\begin{equation}
	\label{eq:det3}
	\det \bS = \det(\bS_1) \det(\bS_2) \frac{\det(\bD_{12})}{\det(\bD_{21})^{*}} \alpha 
\end{equation}
where the last factor reads
\begin{equation}
	\alpha = \frac{\det(\bS_2^\text{ii})^{*} \det(\bD_{12}^{-1} + \bS_2^\text{ie}{\bS_2^\text{ee}}^{-1}
	         \bS_2^\text{ei}\bS_1^\text{ii})}{\det(\bS_1^\text{ii}) \det(\bD_{21}^{-1} + \bS_1^\text{ie}
		 {\bS_1^\text{ee}}^{-1}\bS_1^\text{ei}\bS_2^\text{ii})^{*}}\,.
\end{equation}
This factor can be further evaluated by making use of \eqref{eq:app5} yielding
\begin{subequations}
	\begin{align}
		\bS_1^\text{ie}{\bS_1^\text{ee}}^{-1}\bS_1^\text{ei} &=
						   \bS_1^\text{ii} - {{\bS_1^\text{ii}}^\dagger}^{-1}\\
		\bS_2^\text{ie}{\bS_2^\text{ee}}^{-1}\bS_2^\text{ei} &=
						   \bS_2^\text{ii} - {{\bS_2^\text{ii}}^\dagger}^{ -1}\,.
	\end{align}
\end{subequations}
Employing those expressions and the definitions \eqref{eq:d12} and \eqref{eq:d21}, we find that
\begin{equation}
	\label{eq:alpha}
	\alpha = \frac{\det({\bS_2^\text{ii}}^\dagger - \bS_1^\text{ii})}%
		      {\det(\bS_1^\text{ii} - {\bS_2^\text{ii}}^\dagger)} = (-1)^{n^\text{i}}
\end{equation}
is only a phase factor depending on the number $n^\text{i}$ of internal channels.
Finally, Sylvester's determinant identity implies $\det\bD_{12} = \det\bD_{21}$, so that
we get from \eqref{eq:det3} our first main result
\begin{equation}
	\label{eq:detFinal}
	\det \bS = \det(\bS_1 \star \bS_2)
	= (-1)^{n^\text{i}} \det(\bS_1) \det(\bS_2) \frac{\det(\bD_{21})}{\det(\bD_{21})^{*}}\,.
\end{equation}

\section{Application to the Casimir interaction energy}
\label{sec:casimirEnergy}

\begin{figure}
 \centering
 \includegraphics[width=0.6\textwidth]{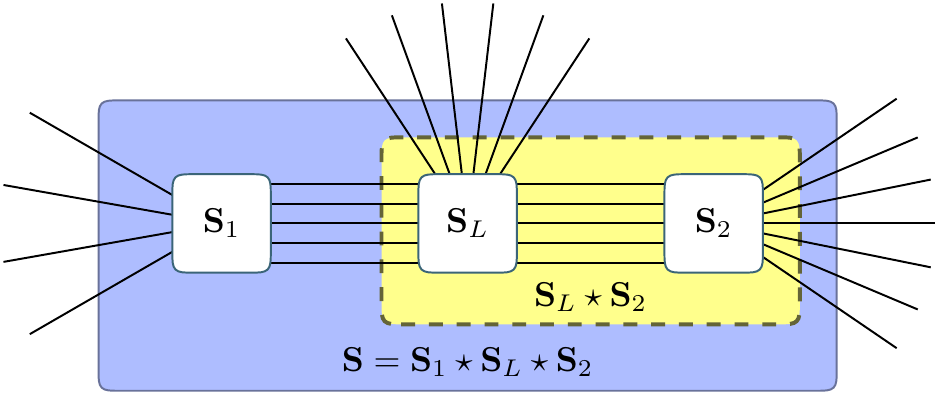}
 \caption{Set-up required to describe the Casimir effect. Apart from the scattering
	matrices $\mathbf{S}_1$ and $\mathbf{S}_2$, a scattering matrix $\mathbf{S}_L$
	describing the translation over a distance $L$ is needed. In addition to
	the internal channels, all scattering matrices couple also to external channels,
	thus allowing to account for dissipation of the objects and the medium in between.
	In a first step, the combination $\mathbf{S}_L\star\mathbf{S}_2$ indicated
	by the dashed box is considered.}
 \label{fig:translation}
\end{figure}

At first sight it might appear that the result \eqref{eq:detFinal} can directly
be applied to the expression for the Casimir energy \eqref{eq:vacuumEnergy}
between two dissipative objects by replacing the general scattering matrices
$\bS_1$ and $\bS_2$ in \eqref{eq:detFinal} by the scattering
matrices $\mathbf{S}_1$ and $\mathbf{S}_2$ of the two dissipative objects.
However, as already pointed out in the first paragraph of
Section~\ref{sec:detFormula}, the translation of the electromagnetic waves
through a potentially dissipative medium between the two objects has not yet
been accounted for.  Actually, we have to consider the set-up depicted in
Figure~\ref{fig:translation}, where in addition to the scattering matrices
$\mathbf{S}_1$ and $\mathbf{S}_2$ a scattering matrix $\mathbf{S}_L$ is
present. This scattering matrix describes the translation of electromagnetic
waves between the bases associated with objects \circled{1} and
\circled{2} over a distance $L$. Concrete examples will be discussed at
the end of this section. Furthermore, $\mathbf{S}_L$ couples to external
channels describing the loss of photons and the influence of noise from the
environment. The global scattering matrix associated with
Figure~\ref{fig:translation} reads
\begin{equation}
	\label{eq:s1sLs2}
	\mathbf{S} = \mathbf{S}_1 \star \mathbf{S}_L \star \mathbf{S}_2\,.
\end{equation}
In the chaining of scattering matrices, we are free to choose the order.  As
indicated by the box marked by a dashed line in Figure~\ref{fig:translation},
we start by evaluating $\mathbf{S}_L \star \mathbf{S}_2$.

With $\mathbf{S}_L$ and $\mathbf{S}_2$ being unitary matrices, we can directly
apply \eqref{eq:detFinal} by replacing $\bS_1$ and $\bS_2$ by
$\mathbf{S}_L$ and $\mathbf{S}_2$, respectively. However, $\bD_{21}$
reflecting the internal round-trips requires some attention. In contrast to
Section~\ref{sec:detFormula}, the internal channels between scattering matrices
$\mathbf{S}_1$ and $\mathbf{S_2}$ are now interrupted by the scattering matrix
$\mathbf{S}_L$ and we should consider as internal only those channels
connecting $\mathbf{S}_2$ and $\mathbf{S}_L$. In contrast, the channels
connecting $\mathbf{S}_1$ and $\mathbf{S}_L$ are to be taken as external for
the present consideration. Since $\mathbf{S}_L$ does not induce backscattering,
it follows that the purely internal part of $\mathbf{S}_L$ vanishes,
$\mathbf{S}^\text{ii}_L=0$. As a consequence, $\bD_{21}$ is a unit
matrix, reflecting the fact that no internal round trips are possible between
$\mathbf{S}_L$ and $\mathbf{S}_2$. From \eqref{eq:detFinal} we then obtain
\begin{equation}
	\label{eq:final2}
	\det(\textbf{S}_{L} \star \textbf{S}_2) = (-1)^{n^\text{i}} \det(\textbf{S}_{L}) \det(\textbf{S}_2)\,.
\end{equation}

In a second step, we apply \eqref{eq:detFinal} with $\bS_1$ and 
$\bS_2$ replaced by $\mathbf{S}_1$ and $\mathbf{S}_L \star \mathbf{S}_2$
and find together with \eqref{eq:final2}
\begin{equation}
   \label{eq:final2sec}
   \det \mathbf{S} = \det(\mathbf{S}_1) \det(\mathbf{S}_2) \det(\mathbf{S}_{L})
	             \frac{\det(\bD_{21})}{\det(\bD_{21})^{*}}\,.
\end{equation}
Apart from $\bS^\text{ii}_1=\mathbf{S}^\text{ii}_1$, the matrix
$\bD_{21}$ contains also the coupling between the internal channels due
to reflection by the chain of scattering matrices $\mathbf{S}_L \star \mathbf{S}_2$.
As explained before, $\mathbf{S}_L$ does not by itself lead to a coupling of
internal channels linked to object \circled{1}. This can happen only by means
of $\mathbf{S}^\text{ii}_2$ sandwiched between translation matrices $\mathbf{T}^\text{ii}_{12}$
and $\mathbf{T}^\text{ii}_{21}$ through a dissipative medium over the distance $L$ from
object \circled{1} to object \circled{2} and back. In the last factor of
\eqref{eq:final2sec} we thus have to set
\begin{equation}
  \label{eq:d21final}
  \bD_{21} = \left(1-\mathbf{S}^\text{ii}_{1}\mathbf{T}^\text{ii}_{12}
	                     \mathbf{S}^\text{ii}_{2}\mathbf{T}^\text{ii}_{21}\right)^{-1}\,.
\end{equation}
We note that in the presence of a dissipative medium, $\mathbf{T}^\text{ii}_{12}$
and $\mathbf{T}^\text{ii}_{21}$ are non-unitary matrices.

We can now insert \eqref{eq:final2sec} together with \eqref{eq:d21final} into 
\eqref{eq:vacuumEnergy} to obtain the change in the vacuum energy due to the
dissipative scatterers separated by a dissipative medium. To obtain the Casimir
interaction energy, we need to identify the part which depends on the distance
$L$ between the two objects. In \eqref{eq:final2sec}, the first two factors depend
only on properties of the individual objects and are thus irrelevant for the
Casimir interaction energy. Only the last two factors depend on $L$. However,
the global scattering matrix $\mathbf{S}$ contains a trivial dependence on $L$
arising from the shift of the basis discussed before \eqref{eq:s1sLs2}. This
effect would survive even in the absence of the objects \circled{1} and \circled{2},
in which case the Casimir interaction energy vanishes. We are thus left with
the last factor. In view of \eqref{eq:vacuumEnergy} and \eqref{eq:scatteringPhase},
we finally obtain for the Casimir interaction energy
\begin{align}
	E_{\text{Cas}}(L) &= \hbar \int_{0}^{\infty} \frac{d \omega}{2 \pi}\, \text{Im}\, \log \det \bD_{21}^{-1} \nonumber \\
	\label{eq:lifshitz}
	&= \hbar \int_{0}^{\infty} \frac{d \omega}{2 \pi}\, \text{Im}\, \log \det
	(\textbf{1}-\textbf{S}_{1}^\text{ii}\textbf{T}_{12}^\text{ii}\textbf{S}_{2}^\text{ii}\textbf{T}_{21}^\text{ii})\,.
\end{align}
This expression depends only on the parts of the scattering matrices pertaining
to the internal channels. Nevertheless, the properties of these parts reflect
the dissipative properties of the objects and the medium in between.

In the form \eqref{eq:lifshitz}, the expression for the Casimir interaction
energy is quite general and basis-independent. The
Dzyaloshinskii-Lifshitz-Pitaevskii formula~\cite{dzyaloshinskii1961general} is
recovered in the case of a plane-plane geometry. In this geometry, it makes
sense to work in a basis of plane waves characterized by the quantum numbers
$\{\omega,\mathbf{q},\varsigma\}$ where $\mathbf{q}=\mathbf{k} -
(\mathbf{k}\cdot\hat{\mathbf{L}})\hat{\mathbf{L}}$ is the transverse part of
the wave vector $\mathbf{k}$ with respect to the unit vector $\hat{\mathbf{L}}$
normal to the two planes~\footnote{Note that $\mathbf{q}$ is a real quantity
since \text{Im}[$\mathbf{k}$] is perpendicular to surfaces of constant
amplitudes.} and $\varsigma$ denotes the polarization. In this basis and this
geometry, both the scattering matrix $\textbf{S}^\text{ii}$ and the translation
matrix $\mathbf{T}^\text{ii}$ are diagonal with matrix elements given by the
Fresnel reflection amplitude $r(\omega,\mathbf{q},\varsigma)$ and
$\exp\left(i(n^{2}\omega^{2}/c^{2}-\mathbf{q}^{2})^{1/2}L\right)$,
respectively. The lossy propagation is conveniently described by introducing a
complex refractive index $n(\omega)$ whose imaginary part is identified with
the attenuation constant.

Another useful basis is the multipole basis
$\{\omega,\ell,m,\varsigma\}$ whenever the system under study has some degree of
spherical symmetry. For a sphere, the scattering matrix $\textbf{S}^\text{ii}$ is
diagonal with elements determined by the Mie scattering amplitudes. The set of
internal channels between a sphere and another object consists of an infinite
number of multipoles arising from translation formulas between spherical waves
(see \emph{e.g.}~\cite{wittmann1988spherical}), so that the translation matrix
$\textbf{T}^\text{ii}$ is not diagonal.

For geometries involving gratings (see \emph{e.g.}~\cite{guerout2013thermal}),
one works once again in a plane-wave basis. In this case, it is the scattering
matrix which is not diagonal due to the non-specular nature of the reflection
by a grating. Therefore, the plane-plane geometry is one of the few examples
where both scattering and translation matrices are diagonal (the other one
being the somewhat unrealistic geometry consisting of two concentric spheres).
In general, at least one of the two matrices is not diagonal. It is possible to
treat in a similar way non-specular scattering for a Drude metal wiht Ohmic
behaviour related to a disordered distribution of impurities
\cite{cherroret2017}.


\section{Conclusion} 
\label{sec:conclusion}

We have derived an expression for the Casimir interaction energy between
dissipative objects embedded in a dissipative medium using the formalism of the
scattering theory. The determinant of the total scattering matrix can be
factored out into parts depending or not on the distance between the objects.
The Casimir interaction energy is expressed using the distance-dependent part.
Our final result \eqref{eq:lifshitz} depends exclusively on scattering matrix
elements involving \emph{internal} channels. Dissipation thus appears only
implicitly in the scattering amplitudes as the blocks over the internal
channels are non-unitary.


\appendix
\section{Useful lemmas} 
\label{sec:appendix}

In this appendix, we gather several relations pertaining to block matrices
which are required in the main part of the text. Let
\begin{equation}
	\label{eq:app1}
	\textbf{M} = \begin{pmatrix}
		\textbf{A} & \textbf{B} \\
		\textbf{C} & \textbf{D}
	\end{pmatrix}
\end{equation}
be a $2 \times 2$ block matrix. Its determinant can expressed either as
\begin{equation}
	\label{eq:app2a}
	\det \textbf{M} = \det(\textbf{A}) \det(\textbf{M}/\textbf{A})
\end{equation}
or
\begin{equation}
	\label{eq:app2}
	\det \textbf{M} = \det(\textbf{D}) \det(\textbf{M}/\textbf{D})
\end{equation}
provided that the blocks $\mathbf{A}$ and/or $\mathbf{D}$ are
invertible. The Schur complements of the blocks $\mathbf{A}$
and $\mathbf{D}$ in $\mathbf{M}$ are defined as
\begin{subequations}
    \begin{align}
	\textbf{M}/\textbf{A} &= \textbf{D} - \textbf{C} \textbf{A}^{-1} \textbf{B}\\
	\label{eq:schur_d}
	\textbf{M}/\textbf{D} &= \textbf{A} - \textbf{B} \textbf{D}^{-1} \textbf{C}\,,
    \end{align}
\end{subequations}
respectively. From the two expressions \eqref{eq:app2a} and \eqref{eq:app2} for the
determinant of $\mathbf{M}$, one obtains the matrix determinant lemma
\begin{equation}
	\label{eq:app6}
	\det(\textbf{A}+\textbf{B}\textbf{D}\textbf{C}) = \det(\textbf{A})\det(\textbf{D})\det(\textbf{D}^{-1} + \textbf{C} \textbf{A}^{-1} \textbf{B})\,,
\end{equation}
if the lower right block in $\mathbf{M}$ is replaced by $-\mathbf{D}^{-1}$.

For the remainder of this section, we assume $\mathbf{M}$ to be unitary. If both
blocks $\textbf{A}$ and $\textbf{D}$ are invertible, the relation
$\textbf{M}^{\dagger} = \textbf{M}^{-1}$ reads
\begin{equation}
	\label{eq:app3}
	\begin{pmatrix}
		\textbf{A}^{\dagger} & \textbf{C}^{\dagger} \\
		\textbf{B}^{\dagger} & \textbf{D}^{\dagger}
	\end{pmatrix}=
	\begin{pmatrix}
		(\textbf{M}/\textbf{D})^{-1} & -(\textbf{M}/\textbf{D})^{-1} \textbf{B} \textbf{D}^{-1} \\
		-\textbf{D}^{-1} \textbf{C} (\textbf{M}/\textbf{D})^{-1} & (\textbf{M}/\textbf{A})^{-1}
	\end{pmatrix}\,.
\end{equation}
From the upper left block together with \eqref{eq:app2}, we find
\begin{equation}
	\label{eq:app4}
	\det \textbf{M} = \frac{\det(\textbf{D})}{\det(\textbf{A}^{\dagger})}\,.
\end{equation}
Together with \eqref{eq:schur_d}, the same relation yields
\begin{equation}
	\label{eq:app5}
	\textbf{B} \textbf{D}^{-1} \textbf{C} = \textbf{A} - \textbf{A}^{\dagger -1}\,.
\end{equation}




\begin{thebibliography}{99}
\begin{small}

\bibitem{casimir1948attraction}
Casimir, H.B.G.
On the attraction between two perfectly conducting plates.
{\em Proc. K. Ned. Akad. Wet} \textbf{1948}, {\em 51},~793 --796.

\bibitem{lifshitz1956theory}
Lifshitz, E.M.
The theory of molecular attractive forces between solids.
{\em Sov. Phys. JETP} \textbf{1956}, {\em 2},~73--83.

\bibitem{schwinger1978casimir}
Schwinger, J.; DeRaad, L.L.; Milton, K.A.
Casimir effect in dielectrics.
{\em Ann. Phys. (N.Y.)} \textbf{1978}, {\em {115}},~{1--23},
doi:10.1016/0003-4916(78)90172-0.

\bibitem{decca2005precise}
Decca, R.S.; L{\'o}pez, D.; Fischbach, E.; Klimchitskaya, G.L.; Krause, D.E.;
Mostepanenko, V.M.
Precise comparison of theory and new experiment for the Casimir force
leads to stronger constraints on thermal quantum effects and long-range
interactions.
{\em Ann. Phys. (N.Y.)} \textbf{2005}, {\em 318},~37--80,
doi:10.1016/j.aop.2005.03.007.

\bibitem{decca2007tests}
Decca, R.S.; L{\'o}pez, D.; Fischbach, E.; Klimchitskaya, G.L.; Krause, D.E.;
Mostepanenko, V.M.
Tests of new physics from precise measurements of the Casimir
pressure between two gold-coated plates.
{\em Phys. Rev. D} \textbf{2007}, {\em 75},~077101,
doi:10.1103/PhysRevD.75.077101.

\bibitem{masuda2009limits}
Masuda, M.; Sasaki, M.
Limits on nonstandard forces in the submicrometer range.
{\em Phys. Rev. Lett.} \textbf{2009}, {\em 102},~171101,
doi:10.1103/PhysRevLett.102.171101.

\bibitem{sushkov2011new}
Sushkov, A.O.; Kim, W.J.; Dalvit, D.A.R.; Lamoreaux, S.K.
New experimental limits on non-Newtonian forces in the micrometer
range.
{\em Phys. Rev. Lett.} \textbf{2011}, {\em 107},~171101,
doi:10.1103/PhysRevLett.107.171101.

\bibitem{chang2012gradient}
Chang, C.-C.; Banishev, A.A.; Castillo-Garza, R.; Klimchitskaya, G.L.;
Mostepanenko, V.M.; Mohideen, U.
Gradient of the Casimir force between Au surfaces of a sphere and a
plate measured using an atomic force microscope in a frequency-shift
technique.
{\em Phys. Rev. B} \textbf{2012}, {\em {85}},~165443,
doi:10.1103/PhysRevB.85.165443.

\bibitem{lambrecht2000casimir}
Lambrecht, A.; Reynaud, S.
Casimir force between metallic mirrors.
{\em Eur. Phys. J. D} \textbf{2000}, {\em {8}},~{309--318},
doi:10.1007/s100530050041.

\bibitem{bostrom2000thermal}
Bostr{\"o}m, M.; Sernelius, B.E.
Thermal effects on the Casimir force in the 0.1--5 $\mu$m range.
{\em Phys. Rev. Lett.} \textbf{2000}, {\em 84},~4757--4760,
doi:10.1103/PhysRevLett.84.4757.

\bibitem{milton2009recent}
Milton, K.A.
Recent developments in the Casimir effect.
{\em J. Phys.: Conf. Ser.} \textbf{2009}, {\em 161},~012001,
doi:10.1088/1742-6596/161/1/012001.

\bibitem{rahi2009scattering}
Rahi, S.J.; Emig, T.; Graham, N.; Jaffe, R.L.; Kardar, M.
Scattering theory approach to electrodynamic Casimir forces.
{\em Phys. Rev. D} \textbf{2009}, {\em 80},~085021,
doi:10.1103/PhysRevD.80.085021.

\bibitem{klimchitskaya2009casimir}
Klimchitskaya, G.L.; Mohideen, U.; Mostepanenko, V.M.
The Casimir force between real materials: Experiment and theory.
{\em Rev. Mod. Phys.} \textbf{2009}, {\em 81},~1827--1885,
doi:10.1103/RevModPhys.81.1827.

\bibitem{ingold2009quantum}
Ingold, G.-L.; Lambrecht, A.; Reynaud, S.
Quantum dissipative Brownian motion and the Casimir effect.
{\em Phys. Rev. E} \textbf{2009}, {\em 80},~041113,
doi:10.1103/PhysRevE.80.041113.

\bibitem{reynaud2017casimir}
Reynaud, S.; Lambrecht, A.
{Casimir} Forces and Vacuum Energy.
Quantum Optics and Nanophotonics; Fabre, C.; Sandoghdar, V.; Treps,
N.; Cugliandolo, L.F., Eds. Oxford University Press,  2017, Lecture Notes of
the Les Houches Summer School 101, pp. 407--455,
doi:10.1093/oso/9780198768609.003.0009.

\bibitem{speake2003forces}
Speake, C.C.; Trenkel, C.
Forces between conducting surfaces due to spatial variations of
surface potential.
{\em Phys. Rev. Lett.} \textbf{2003}, {\em 90},~160403,
doi:10.1103/PhysRevLett.90.160403.

\bibitem{kim2010surface}
Kim, W.J.; Sushkov, A.O.; Dalvit, D.A.R.; Lamoreaux, S.K.
Surface contact potential patches and Casimir force measurements.
{\em Phys. Rev. A} \textbf{2010}, {\em 81},~022505,
doi:10.1103/PhysRevA.81.022505.

\bibitem{behunin2012modeling}
Behunin, R.O.; Intravaia, F.; Dalvit, D.A.R.; Maia~Neto, P.A.; Reynaud, S.
Modeling electrostatic patch effects in Casimir force measurements.
{\em Phys. Rev. A} \textbf{2012}, {\em 85},~012504,
doi:10.1103/PhysRevA.85.012504.

\bibitem{behunin2012electrostatic}
Behunin, R.O.; Zeng, Y.; Dalvit, D.A.R.; Reynaud, S.
Electrostatic patch effects in Casimir-force experiments performed in
the sphere-plane geometry.
{\em Phys. Rev. A} \textbf{2012}, {\em 86},~052509,
doi:10.1103/PhysRevA.86.052509.

\bibitem{behunin2014kelvin}
Behunin, R.O.; Dalvit, D.A.R.; Decca, R.S.; Genet, C.; Jung, I.W.; Lambrecht,
A.; Liscio, A.; L{\'o}pez, D.; Reynaud, S.; Schnoering, G.; Voisin, G.; Zeng, Y.
Kelvin probe force microscopy of metallic surfaces used in Casimir force measurements.
{\em Phys. Rev. A} \textbf{2014}, {\em {90}},~{062115},
doi:10.1103/PhysRevA.90.062115.

\bibitem{intravaia2009casimir}
Intravaia, F.; Henkel, C.
Casimir interaction from magnetically coupled eddy currents.
{\em Phys. Rev. Lett.} \textbf{2009}, {\em 103},~130405,
doi:10.1103/PhysRevLett.103.130405.

\bibitem{guerout2014derivation}
Gu{\'e}rout, R.; Lambrecht, A.; Milton, K.A.; Reynaud, S.
Derivation of the Lifshitz-Matsubara sum formula for the Casimir
pressure between metallic plane mirrors.
{\em Phys. Rev. E} \textbf{2014}, {\em 90},~042125,
doi:10.1103/PhysRevE.90.042125.

\bibitem{guerout2016lifshitz}
Gu{\'e}rout, R.; Lambrecht, A.; Milton, K.A.; Reynaud, S.
Lifshitz-Matsubara sum formula for the Casimir pressure between
magnetic metallic mirrors.
{\em Phys. Rev. E} \textbf{2016}, {\em 93},~022108,
doi:10.1103/PhysRevE.93.022108.

\bibitem{bordag2011drude}
Bordag, M.
Drude model and Lifshitz formula.
{\em Eur. Phys. J. C} \textbf{2011}, {\em {71}},~1788,
doi:10.1140/epjc/s10052-011-1788-x.

\bibitem{jancovici2005casimir}
Jancovici, B.; {\v S}amaj, L.
Casimir force between two ideal-conductor walls revisited.
{\em Europhys. Lett.} \textbf{2005}, {\em 72},~35--41,
doi:10.1209/epl/i2005-10201-5.

\bibitem{buenzli2005casimir}
Buenzli, P.R.; Martin, P.A.
The Casimir force at high temperature.
{\em Europhys. Lett.} \textbf{2005}, {\em 72},~42--48,
doi:10.1209/epl/i2005-10200-6.

\bibitem{intravaia2012casimir}
Intravaia, F.; Behunin, R.
Casimir effect as a sum over modes in dissipative systems.
{\em Phys. Rev. A} \textbf{2012}, {\em 86},~062517,
doi:10.1103/PhysRevA.86.062517.

\bibitem{bordag2017casimir}
Bordag, M.
Casimir and Casimir-Polder forces with dissipation from first
principles.
{\em Phys. Rev. A} \textbf{2017}, {\em 96},~062504,
doi:10.1103/PhysRevA.96.062504.

\bibitem{jaekel1991casimir}
Jaekel, M.T.; Reynaud, S.
Casimir force between partially transmitting mirrors.
{\em J. Phys.~I France} \textbf{1991}, {\em 1},~1395--1409,
doi:10.1051/jp1:1991216.

\bibitem{ingold2015casimir}
Ingold, G.-L.; Lambrecht, A.
Casimir effect from a scattering approach.
{\em Am. J. Phys.} \textbf{2015}, {\em 83},~156--162,
doi:10.1119/1.4896197.

\bibitem{genet2003casimir}
Genet, C.; Lambrecht, A.; Reynaud, S.
Casimir force and the quantum theory of lossy optical cavities.
{\em Phys. Rev. A} \textbf{2003}, {\em 67},~043811,
doi:10.1103/PhysRevA.67.043811.

\bibitem{lambrecht2006casimir}
Lambrecht, A.; Maia~Neto, P.A.; Reynaud, S.
The Casimir effect within scattering theory.
{\em New J. Phys.} \textbf{2006}, {\em 8},~243,
doi:10.1088/1367-2630/8/10/243.

\bibitem{engquist1981definition}
Engquist, H.L.; Anderson, P.W.
Definition and measurement of the electrical and thermal resistances.
{\em Phys. Rev. B} \textbf{1981}, {\em {24}},~{1151--1154},
doi:10.1103/PhysRevB.24.1151.

\bibitem{huttner1992quantization}
Huttner, B.; Barnett, S.M.
Quantization of the electromagnetic-field in dielectrics.
{\em Phys. Rev. A} \textbf{1992}, {\em {46}},~{4306--4322},
doi:10.1209/0295-5075/18/6/003.

\bibitem{matloob1995electromagnetic}
Matloob, R.; Loudon, R.; Barnett, S.M.; Jeffers, J.
Electromagnetic field quantization in absorbing dielectrics.
{\em Phys. Rev. A} \textbf{1995}, {\em {52}},~{4823--4838},
doi:10.1103/PhysRevA.52.4823.

\bibitem{schwinger1954theory}
Schwinger, J.
The theory of quantized fields. VI.
{\em Phys. Rev.} \textbf{1954}, {\em {94}},~{1362--1384},
doi:10.1103/PhysRev.94.1362.

\bibitem{balian1977electromagnetic}
Balian, R.; Duplantier, B.
Electromagnetic waves near perfect conductors. I. Multiple scattering
expansions. Distribution of modes.
{\em Ann. Phys. (N.Y.)} \textbf{1977}, {\em 104},~300--335,
doi:10.1016/0003-4916(77)90334-7.

\bibitem{plunien1986casimir}
Plunien, G.; M{\"u}ller, B.; Greiner, W.
The Casimir effect.
{\em Phys. Rep.} \textbf{1986}, {\em {134}},~{87--193},
doi:10.1016/0370-1573(86)90020-7.

\bibitem{souma2002local}
Souma, S.; Suzuki, A.
Local density of states and scattering matrix in
quasi-one-dimensional systems.
{\em Phys. Rev. B} \textbf{2002}, {\em 65},~115307,
doi:10.1103/PhysRevB.65.115307.

\bibitem{dzyaloshinskii1961general}
Dzyaloshinskii, I.E.; Lifshitz, E.M.; Pitaevskii, L.P.
General theory of van der Waals' forces.
{\em Sov. Phys. Usp.} \textbf{1961}, {\em 4},~153--176,
doi:10.1070/PU1961v004n02ABEH003330.

\bibitem{wittmann1988spherical}
Wittmann, R.C.
Spherical wave operators and the translation formulas.
{\em IEEE Trans. Antennas Propag.} \textbf{1988}, {\em 36},~1078--1087,
doi:10.1109/8.7220.

\bibitem{guerout2013thermal}
Gu{\'e}rout, R.; Lussange, J.; Chan, H.B.; Lambrecht, A.; Reynaud, S.
Thermal Casimir force between nanostructured surfaces.
{\em Phys. Rev. A} \textbf{2013}, {\em 87},~052514,
doi:10.1103/PhysRevA.87.052514.

\bibitem{cherroret2017}
Cherroret, N; Cr{\'e}pin, P.-P.; Gu{\'e}rout, R.; Lambrecht, A.
Casimir-Polder force fluctuations as spatial probes of dissipation 
in metals.
{\em EPL} \textbf{2017}, {\em 117},~63001,
doi:10.1209/0295-5075/117/63001.
\end{small}

\end{thebibliography}
\end{document}